\documentclass[12pt,a4paper,dvipdfmx]{article}
\usepackage{graphicx}
\usepackage{amssymb,amsfonts,amsmath}
\usepackage{cite}
\usepackage{bm}

\usepackage{color}
\usepackage{multirow}

\title{A Gillespie algorithm for non-Markovian stochastic processes}
\bigskip
\author{Naoki Masuda${}^{1,*}$ and Luis E. C. Rocha${}^{2,3}$
\ \\
\ \\
${}^{1}$
Department of Engineering Mathematics, University of Bristol, Bristol, UK.\\
${}^{2}$
Department of Mathematics and naXys, Universit\'e de Namur, Namur, Belgium.\\
${}^{3}$
Department of Public Health Sciences, Karolinska Institutet, Stockholm, Sweden.\\
\ \\
$^*$ Corresponding author (naoki.masuda@bristol.ac.uk)}

\begin{document}
\setlength{\baselineskip}{24pt}

\maketitle

\section*{Abstract}
The Gillespie algorithm provides statistically exact methods for simulating stochastic dynamics modelled as interacting sequences of discrete events including systems of biochemical reactions or earthquake occurrences, networks of queuing processes or spiking neurons, and epidemic and opinion formation processes on social networks. Empirically, the inter-event times of various phenomena obey long-tailed distributions. The Gillespie algorithm and its variants either assume Poisson processes (i.e., exponentially distributed inter-event times), use particular functions for time courses of the event rate, or work for non-Poissonian renewal processes, including the case of long-tailed distributions of inter-event times, but at a high computational cost. In the present study, we propose an innovative Gillespie algorithm for renewal processes on the basis of the Laplace transform. The algorithm makes use of the fact that a class of point processes is represented as a mixture of Poisson processes with different event rates. The method is applicable to multivariate renewal processes whose survival function of inter-event times is completely monotone. It is an exact algorithm and works faster than a recently proposed Gillespie algorithm for general renewal processes, which is exact only in the limit of infinitely many processes. We also propose a method to generate sequences of event times with a tunable amount of positive correlation between inter-event times. We demonstrate our algorithm with exact simulations of epidemic processes on networks, finding that a realistic amount of positive correlation in inter-event times only slightly affects the epidemic dynamics.

\section{Introduction}\label{sec:introduction}

Various complex systems are driven by interactions between subsystems via time-stamped discrete events. For example, in chemical systems, a chemical reaction event changes the number of reagents of particular types in a discrete manner. Stochastic point processes, in particular Poisson processes assuming that events occur independently and at a constant rate over time, are a central tool for emulating the dynamics of chemical systems \cite{Vankampen2007book}. They are also useful in simulating epidemic processes in a population \cite{Daley2002book} and many other systems.

Consider an event-driven system in which events are generated by Poisson processes running in parallel.
In chemical reaction systems, each Poisson process (possibly with different rates for each process) is attached to one reaction. In epidemic processes taking place on human or animal contact networks (i.e., graphs), each Poisson process is assigned to an individual or a link, which may potentially transmit the infection. The event rate of some of the Poisson processes may change with the occurrence of a reaction or infection within the entire system. The simplest simulation method is to discretise time and then judge whether or not an event occurs in each time window, for individual processes. This widely used method is suboptimal because the size of the time window must be sufficiently small to obtain high accuracy, which is computationally expensive \cite{Vestergaard2015PlosComputBiol}. The Gillespie algorithm is an efficient and statistically exact algorithm for multivariate Poisson processes \cite{Kendall1950JRStatSocSerB,Gillespie1976JComputPhys,Gillespie1977JPhysChem}. The Gillespie algorithm, or, in particular, the direct method of Gillespie \cite{Gillespie1976JComputPhys,Gillespie1977JPhysChem}, exploits the fact that a superposition of independent Poisson processes is a single Poisson process whose event rate is the sum of those of the constituent Poisson processes. Using this mathematical property, only a single Poisson process needs to be emulated in the Gillespie algorithm.

However, for various real-world systems in which multivariate point processes have been applied (both with and without network structure), event sequences are clearly non-Poissonian. In particular, inter-event times typically obey long-tailed distributions, which are inconsistent with the exponential (i.e., short-tailed) distribution that Poisson processes produce. Examples of long-tailed distributions of inter-event times include earthquake occurrences \cite{Bak2002PhysRevLett-earthquake,Corral2004PhysRevLett}, firing of neurons \cite{Softky1993JNeurosci,Baddeley1997ProcRSocLondB}, social networks and the Internet \cite{Barabasi2005Nature,VazquezA2006PhysRevE-burst,GohBarabasi2008EPL,HolmeSaramaki2012PhysRep}, financial transactions \cite{Raberto2002PhysicaA,Masoliver2003PhysRevE}, and crimes \cite{Mohler2011JAmStatAssoc,Johnson2012Policing}. Therefore, multivariate point processes that are not necessarily Poissonian have been used in these applications, e.g., triggered seismicity \cite{Ogata1999PureApplGeophys,Helmstetter2002JGeophysRes}, networks of spiking neurons \cite{Truccolo2005JNeurophysiol,Pernice2011PlosComputBiol}, epidemic processes
\cite{HolmeSaramaki2012PhysRep,Masuda2013F1000}, opinion formation models \cite{Fernandezgracia2011PhysRevE,Takaguchi2011PhysRevE}, finance \cite{Bowsher2007JEconometr,Filimonov2012PhysRevE}, and criminology \cite{Mohler2011JAmStatAssoc,Porter2012AnnApplStat}. These applications call for numerical methods to efficiently and accurately simulate interacting and non-Markovian point processes.

A reasonable description of event sequences in these phenomena requires, at the very least, renewal processes, in which inter-event times are independently generated from a given distribution \cite{Cox1962book,Feller1971book2}. Along these lines, one numerical approach is to use the modified next reaction method \cite{Anderson2007JChemPhys}, which improves on both the so-called Gillespie's first reaction method \cite{Gillespie1976JComputPhys} and the next reaction method \cite{GibsonBruck2000JPhysChemA}. The basic idea behind these methods is to draw the next event time for all processes from predetermined distributions, select the process that has generated the minimum waiting time to the next event, execute the event, and repeat. However, for non-Poissonian renewal processes, it is generally difficult to numerically solve the equations that determine the next event time although these methods call for the generation of only half as many random numbers in comparison to the Gillespie algorithm \cite{Vestergaard2015PlosComputBiol}. In addition, we can easily halve the number of random variates required in the Gillespie algorithm, such that the next reaction method and the Gillespie algorithm demand the same number of random variates (Appendix~A). In what follows, we restrict ourselves to the Gillespie algorithm and its variants.
 
Motivated by applications to chemical reactions, many extensions of the Gillespie algorithm in the case of non-Poissonian processes assume that the dynamical change in the event rate is exogenously driven in particular functional forms \cite{LuVolfson2004SystBiol,Carletti2012ComputMathMethodsMed}. These extensions are not applicable for general renewal processes, because the event rate of a constituent process is a function of the time since the last event of the process, which depends on that process. In other words, we cannot assume a common exogenous driver. Bogu\~{n}\'{a} and colleagues extended the Gillespie algorithm to be applicable for general renewal processes
\cite{BogunaLafuerza2014PhysRevE} (also see \cite{Vestergaard2015PlosComputBiol} for further developments). However, the algorithm has practical limitations. First, it is not accurate when the number of ongoing renewal processes is small \cite{BogunaLafuerza2014PhysRevE}. This can result in a considerable amount of approximation error in the beginning or final stages of the dynamics of epidemic or opinion formation models, for example, in which only a small number of processes is active, even in a large population \cite{Vestergaard2015PlosComputBiol}. Second, it is necessary to recalculate the instantaneous event rate of each process following the occurrence of every event in the entire population, a procedure that can be computationally expensive.

In the present study, we propose an innovative Gillespie algorithm, the Laplace Gillespie algorithm, which is applicable to non-Poissonian renewal processes. It exploits the mathematical properties of the Laplace transform, is accurate for an arbitrary number of ongoing renewal processes, and runs faster than the previous algorithm \cite{BogunaLafuerza2014PhysRevE}. This article is organised as follows. In section~\ref{sec:Gillespie algorithm}, we review the original Gillespie algorithm for Poisson processes. In section~\ref{sec:nMGA}, we review the previous extension of the Gillespie algorithm to general renewal processes \cite{BogunaLafuerza2014PhysRevE}. In section~\ref{sec:LGA}, we introduce the Laplace Gillespie algorithm, together with theoretical underpinnings and examples. In section~\ref{sec:numerical}, we numerically compare the previous algorithm \cite{BogunaLafuerza2014PhysRevE} and the Laplace Gillespie algorithm. In section~\ref{sub:positive correlation}, we introduce a method to generate event sequences with positive correlation in inter-event times, as is typically observed in human behaviour and natural phenomena \cite{GohBarabasi2008EPL}. In section~\ref{sec:epidemic}, we demonstrate our methods by performing exact simulations of an epidemic process in which inter-event times follow a power-law distribution. In section~\ref{sec:discussion}, we discuss the results, including limitations of the proposed algorithm. The codes used in our numerical simulations are available in the Supplementary Materials.

\section{Gillespie algorithm\label{sec:Gillespie algorithm}}

The original Gillespie algorithm \cite{Kendall1950JRStatSocSerB,Gillespie1976JComputPhys,Gillespie1977JPhysChem} assumes $N$ independent Poisson processes with a rate of $\lambda_i$ ($1\le i\le N$) running in parallel. Because of the independence of the different Poisson processes, the superposition of the $N$ processes is a Poisson process with rate $\sum_{i=1}^N \lambda_i$. Therefore, we first draw $\Delta t$, an increment in time to the next event of the superposed Poisson process, from the exponential distribution given by
\begin{equation}
\phi(\Delta t) = \left(\sum_{i=1}^N \lambda_i\right)
e^{-\left(\sum_{i=1}^N \lambda_i\right)\Delta t}.
\label{eq:phi(Delta t) Gillespie}
\end{equation}
Because the survival function (i.e., the probability that a random variable is larger than a certain value) of $\phi(\Delta t)$ is given by $\int_{\Delta t}^{\infty} \phi(t^{\prime}){\rm d}t^{\prime} = e^{-\left(\sum_{i=1}^N \lambda_i\right)\Delta t}$, we obtain $\Delta t = -\log u\big/ \left(\sum_{i=1}^N \lambda_i\right)$, where $u$ is a random variate drawn from the uniform density on the interval $[0, 1]$. Then, we determine the process $i$ that has produced the event with probability
\begin{equation}
\Pi_i = \frac{\lambda_i}{\sum_{i=1}^N \lambda_i}.
\label{eq:Pi(i) Gillespie}
\end{equation}
Finally, we advance the time by $\Delta t$ and repeat the procedure. Following the occurrence of an event, any $\lambda_i$ is permitted to change.

\section{Non-Markovian Gillespie algorithm\label{sec:nMGA}} 

Now consider $N$ renewal processes running in parallel, and denote by $\psi_i(\tau)$ the probability density function of inter-event times for the $i$th process ($1\le i\le N$). If the process is Poissonian, we obtain $\psi_i(\tau) = \lambda_i e^{-\lambda_i \tau}$. For such a population of general renewal processes, Bogu\~{n}\'{a} and colleagues proposed an extension of the Gillespie algorithm, which they called the non-Markovian Gillespie algorithm (nMGA) \cite{BogunaLafuerza2014PhysRevE}.

To begin with, we explain their exact Gillespie algorithm for general renewal processes, which is the basis of the nMGA. A short derivation of the exact Gillespie algorithm is given in Appendix~B. We denote by $t_i$ the time elapsed since the last event of the $i$th process, and by
\begin{equation}
\Psi_i(t_i) = \int_{t_i}^\infty \psi_i(\tau^{\prime}){\rm d}\tau^{\prime}
\label{eq:Psi_i(t_i)}
\end{equation}
the survival function of the $i$th process (i.e., the probability that the inter-event time is larger than $t_i$). We also set
\begin{equation}
\Phi(\Delta t | \{t_j\}) = \prod_{j=1}^N \frac{\Psi_j(t_j+\Delta t)}{\Psi_j(t_j)},
\label{eq:Phi(Delta t | set of t_j) Gillespie}
\end{equation}
which is in fact equal to the probability that no process generates an event for time $\Delta t$ (see Appendix~B). The exact Gillespie algorithm for general renewal processes is given as follows:
\begin{enumerate}

\item Initialise $t_j$ ($1\le j\le N$) for all $j$ (for example, $t_j=0$).

\item Draw the time until the next event, $\Delta t$, by solving $\Phi(\Delta t | \{t_j\}) = u$, where $u$ is a random variate uniformly distributed over $[0, 1]$.

\item Select the process $i$ that has generated the event with probability 
\begin{equation}
\Pi_i \equiv \frac{\lambda_i(t_i+\Delta t)}{\sum_{j=1}^N \lambda_j(t_j+\Delta t)},
\label{eq:Pi(i | Delta t, set of t_j) Gillespie}
\end{equation}
where $\lambda_i(t_i+\Delta t) \equiv \psi_i(t_i+\Delta t) / \Psi_i(t_i+\Delta t)$ is equal to the instantaneous rate of the $i$th process.

\item Update the time since the last event, $t_j$, to $t_j + \Delta t$ ($j\neq i$) and set $t_i=0$.

\item Repeat steps 2--4.

\end{enumerate}

Although this algorithm is statistically exact, step 2 can be time-consuming \cite{BogunaLafuerza2014PhysRevE,Vestergaard2015PlosComputBiol}. To improve performance, Bogu\~{n}\'{a} and colleagues introduced the nMGA. The nMGA is an approximation to the aforementioned algorithm and is exact as $N\to\infty$. When $\Delta t$ is small, as is the case when $N$ is large, Eq.~\eqref{eq:Phi(Delta t | set of t_j) Gillespie} is approximated as
\begin{align}
\Phi(\Delta t | \{t_j\}) =& \exp \left[ -\sum_{j=1}^N \ln\frac{\Psi_j(t_j)}{\Psi_j(t_j+\Delta t)}\right]\notag\\
=& \exp \left[ -\sum_{j=1}^N \ln\frac{\Psi_j(t_j)}{\Psi_j(t_j) - \psi_j(t_j)\Delta t + O(\Delta t^2)}\right]\notag\\
\approx& \exp \left[-\Delta t \left(\sum_{j=1}^N \lambda_j(t_j)\right)\right].
\end{align}
With this approximation, the time until the next event is determined by $\Phi(\Delta t | \{t_j\})$ $\approx$ $\exp \left[-\Delta t \left(\sum_{j=1}^N \lambda_j(t_j)\right)\right]$ $= u$, i.e., $\Delta t = -\ln u\big/ \left(\sum_{j=1}^N \lambda_j(t_j)\right)$. The process that generates the event is determined by setting $\Delta t=0$ in Eq.~\eqref{eq:Pi(i | Delta t, set of t_j) Gillespie}. For a Poisson process, we set $\lambda_i(t_i) = \lambda_i$ to recover the original Gillespie algorithm (cf. Eqs.~\eqref{eq:phi(Delta t) Gillespie} and \eqref{eq:Pi(i) Gillespie}).

\section{Laplace Gillespie algorithm}\label{sec:LGA}

\subsection{Algorithm}

In the nMGA, we update the instantaneous event rates for all the processes $\lambda_j(t_j)$ $(1\le j\le N)$ and their sum following the occurrence of each event. This is because $t_j$ ($1\le j\le N$) is updated following an event. This procedure is time-consuming when $N$ is large; we have to update $\lambda_j(t_j)$ even if the probability density of the inter-event times for the $j$th process is not perturbed by an event that has occurred elsewhere.

To construct an efficient Gillespie algorithm for non-Markovian point processes, we start by considering the following renewal process, called the event-modulated Poisson process. When an event occurs, we first draw the rate of the Poisson process, denoted by $\lambda$, according to a fixed probability density function $p(\lambda)$. Then, we draw the time until the next event according to the Poisson process with rate $\lambda$. Upon the occurrence of the next event, we renew the rate $\lambda$ by redrawing it from $p(\lambda)$. We then repeat these steps.

The event-modulated Poisson process is a mixture of Poisson processes of different rates. It is, in general, a non-Poissonian renewal process and is slightly different from a mixed Poisson process, in which a single rate is initially drawn from a random ensemble and used throughout a realisation \cite{Karr1991book,Grandell1997book}. It is also different from a doubly stochastic Poisson process (also known as a Cox process), in which the rate of the Poisson process is a stochastic process \cite{Kingman1964MathprocCambPhilSoc,Karr1991book,Grandell1997book}, or its subclass called the Markov-modulated Poisson process, in which the event rate switches in time according to a Markov process \cite{Fischer1992PerfEval}. In these processes, the dynamics of the event rate are independent of the occurrence of events. In contrast, for event-modulated Poisson processes, the event rate changes upon the occurrence of events.

An event-modulated Poisson process is a Poisson process when conditioned on the current value of $\lambda$. Therefore, when we simulate $N$ event-modulated Poisson processes, they are independent of each other and of the past event sequences if we are given the instantaneous rate of the $i$th process, denoted by $\lambda_i$ for all $i$ ($1\le i\le N$). This property enables us to construct a Gillespie algorithm. By engineering $p(\lambda)$, we can emulate a range of renewal processes with different distributions of inter-event times. The new Gillespie algorithm, which we call the Laplace Gillespie algorithm (the reason for this name will be made clear in section~\ref{sub:theory}, where we discuss the algorithm's theoretical basis in the Laplace transform), is defined as the Gillespie algorithm for event-modulated Poisson processes. We denote the density of the event rate for the $i$th process by $p_i(\lambda_i)$. The Laplace Gillespie algorithm proceeds as follows:

\begin{enumerate}

\item Initialise each of the $N$ processes by drawing the rate $\lambda_i$ ($1\le i\le N$) according to its density function $p_i(\lambda_i)$.

\item Draw the time until the next event $\Delta t = -\ln u/\sum_{j=1}^N \lambda_j$, where $u$ is a random variate uniformly distributed over $[0, 1]$.

\item Select the process $i$ that has generated the event with probability $\lambda_i/\sum_{j=1}^N \lambda_j$.

\item Draw a new rate $\lambda_i$ according to $p_i(\lambda_i)$. If there are processes $j$ ($1\le j\le N$) for which the statistics of inter-event times have changed following the occurrence of the event generated in steps 2 and 3 (e.g., a decrease in the rate of being infected owing to the recovery of an infected neighbour), modify $p_j(\lambda_j)$ accordingly and draw a new rate $\lambda_j$ from the modified $p_j(\lambda_j)$. The event rates of the remaining processes remain unchanged.

\item Repeat steps 2--4.

\end{enumerate}

\subsection{Theory\label{sub:theory}}

An event-modulated Poisson process is a renewal process. The renewal process is fully characterised by the probability density of inter-event times, $\psi(\tau)$. For an event-modulated Poisson process with probability density of the event rate $p(\lambda)$, we obtain
\begin{equation}
\psi(\tau) = \int_0^{\infty} p(\lambda) \lambda e^{-\lambda \tau} {\rm d}\lambda.
\label{eq:psi(tau) mix}
\end{equation}
Integration of both sides of Eq.~\eqref{eq:psi(tau) mix} yields the survival function of the inter-event times as follows:
\begin{equation}
\Psi(\tau) = \int_{\tau}^{\infty} \psi(\tau^{\prime}){\rm d}\tau^{\prime} = \int_0^{\infty} p(\lambda) e^{-\lambda \tau} {\rm d}\lambda.
\label{eq:Psi(tau) mix}
\end{equation}
Equation~\eqref{eq:Psi(tau) mix} indicates that $\Psi(\tau)$ is the Laplace transform of $p(\lambda)$. Therefore, the necessary and sufficient condition for a renewal process to be simulated by the Laplace Gillespie algorithm is that $\Psi(\tau)$ is the Laplace transform of a probability density function of a random variable taking nonnegative values. Although this statement can be made more rigorous if we replace $p(\lambda){\rm d}\lambda$ by the probability distribution function, we will use the probability density representation for simplicity.

A necessary and sufficient condition for the existence of $p(\lambda)$ is that $\Psi(\tau)$ is completely monotone and $\Psi(0)=1$ \cite{Feller1971book2}. The complete monotonicity is defined by
\begin{equation}
(-1)^n \frac{{\rm d}^n\Psi(\tau)}{{\rm d}\tau^n} \ge 0\quad (\tau\ge 0, n= 0, 1, \ldots).
\label{eq:completely monotone}
\end{equation}
Condition $\Psi(0)=1$ is satisfied by any survival function. With $n=0$, Eq.~\eqref{eq:completely monotone} reads $\Psi(\tau)\ge 0$, which all survival functions satisfy. With $n=1$, Eq.~\eqref{eq:completely monotone} reads $\psi(\tau)\ge 0$, which is also always satisfied. However, Eq.~\eqref{eq:completely monotone} with $n\ge 2$ imposes nontrivial conditions.

\subsection{Examples\label{sub:examples}}

In this section, we give examples of distributions of inter-event times $\psi(\tau)$ for which the Laplace Gillespie algorithm can be used. These examples are summarised in Table~\ref{tab:examples}.

\begin{itemize}

\item Exponential distribution

A Poisson process with rate $\lambda_0$, i.e., $\psi(\tau) = \lambda_0 e^{-\lambda_0 \tau}$, is trivially generated by $p(\lambda)=\delta(\lambda-\lambda_0)$, where $\delta$ is the delta function.

\item Power-law distribution derived from a gamma distribution of $\lambda$

Consider the case in which $p(\lambda)$ is the gamma distribution given by
\begin{equation}
p(\lambda) = \frac{\lambda^{\alpha-1}e^{-\lambda/\kappa}}{\Gamma(\alpha)\kappa^{\alpha}},
\label{eq:p(lambda) gamma}
\end{equation}
where $\Gamma(\alpha)$ is the gamma function, $\alpha$ is the shape parameter, and $\kappa$ is the scale parameter. By combining Eqs.~\eqref{eq:Psi(tau) mix} and \eqref{eq:p(lambda) gamma}, we obtain
\begin{equation}
\Psi(\tau) = \frac{1}{(1+\kappa\tau)^{\alpha}}.
\label{eq:Psi(tau) power law}
\end{equation}
The probability density of inter-event times is given by the following power-law distribution:
\begin{equation}
\psi(\tau) = \frac{\kappa\alpha}{(1+\kappa\tau)^{\alpha+1}}.
\label{eq:psi(tau) power law}
\end{equation}
When $\alpha=1$, $p(\lambda)$ is the exponential distribution and $\psi(\tau)\propto \tau^{-2}$ \cite{Hidalgo2006PhysicaA}. The same mathematical relationship connecting the gamma distribution and a power-law distribution has been used in superstatistics in statistical mechanics \cite{BeckCohen2003PhysicaA} and in a reinforcement learning model for generating discount rates that decay with time according to a power law \cite{Kurthnelson2009PlosOne}.

\item Power-law distribution with an exponential tail derived from a gamma distribution of $\lambda$

Consider a shifted gamma distribution \cite{Grandell1997book} given by
\begin{equation}
p(\lambda) = \begin{cases}
\frac{(\lambda-\lambda_0)^{\alpha-1}e^{-(\lambda-\lambda_0)/\kappa}}{\Gamma(\alpha)} & (\lambda\ge\lambda_0),\\
0 & (0 < \lambda < \lambda_0),
\end{cases}
\label{eq:p(lambda) shifted gamma}
\end{equation}
where $\lambda_0$ is a constant. By combining Eqs.~\eqref{eq:Psi(tau) mix} and \eqref{eq:p(lambda) shifted gamma}, we obtain
\begin{equation}
\Psi(\tau) = \frac{e^{-\lambda_0 \tau}}{(1+\kappa\tau)^{\alpha}}.
\label{eq:Psi(tau) from shifted gamma}
\end{equation}
By differentiating Eq.~\eqref{eq:Psi(tau) from shifted gamma}, we obtain a power-law distribution with an exponential tail given by
\begin{equation}
\psi(\tau) = \frac{e^{-\lambda_0 \tau}}{(1+\kappa\tau)^{\alpha}}\left(\lambda_0+\frac{\kappa\alpha}{1+\kappa\tau}\right).
\end{equation}

\item Power-law distribution derived from a uniform distribution of $\lambda$

Assume that $p(\lambda)$ is a uniform density on $[\lambda_{\min}, \lambda_{\max}]$ \cite{Hidalgo2006PhysicaA}. We obtain
\begin{equation}
\Psi(\tau) = \frac{e^{-\lambda_{\min}\tau}-e^{-\lambda_{\max}\tau}}{\tau\left(\lambda_{\max}-\lambda_{\min}\right)}
\end{equation}
and
\begin{equation}
\psi(\tau) = \frac{\lambda_{\min}e^{-\lambda_{\min}\tau}-\lambda_{\max}e^{-\lambda_{\max}\tau}}{\left(\lambda_{\max}-\lambda_{\min}\right)\tau} + \frac{e^{-\lambda_{\min}\tau}-e^{-\lambda_{\max}\tau}}{\left(\lambda_{\max}-\lambda_{\min}\right)\tau^2}.
\end{equation}

Now suppose that $\lambda_{\min}\ll \lambda_{\max}$. If $\lambda_{\min}>0$, we obtain $\psi(\tau)\propto e^{-\lambda_{\min}\tau}/\tau$ as $\tau\to\infty$, a power-law distribution with an exponential cutoff. If $\lambda_{\min}=0$, we obtain $\psi(\tau)\propto 1/\tau^2$ as $\tau\to\infty$.

\item Power-law distribution derived from a power-law distribution of $\lambda$

Consider the distribution
\begin{equation}
p(\lambda) = (\alpha+1) \lambda^{\alpha},
\end{equation}
where $\alpha>-1$ and $0\le \lambda\le 1$ \cite{Hidalgo2006PhysicaA}. We obtain
\begin{equation}
\Psi(\tau) = \frac{\alpha+1}{\tau^{\alpha+1}} \gamma(\alpha+1,\tau),
\end{equation}
where $\gamma(\alpha+1,\tau) = \int_0^{\tau} x^{\alpha} e^{-x}{\rm d}x$ is the incomplete gamma function, and
\begin{equation}
\psi(\tau) = \frac{\alpha+1}{\tau^{\alpha+2}} \gamma(\alpha+2,\tau).
\end{equation}
As $\tau\to\infty$, we obtain $\psi(\tau)\approx (\alpha+1)\Gamma(\alpha+2)/\tau^{\alpha+2}$.

\item Weibull distribution

The Weibull distribution is defined by
\begin{equation}
\Psi(\tau) = e^{-(\mu\tau)^{\alpha}},
\end{equation}
which yields
\begin{equation}
\psi(\tau) = \alpha\mu^{\alpha}\tau^{\alpha-1} e^{-(\mu\tau)^{\alpha}}.
\end{equation}
The Weibull distribution with $\alpha=1$ is an exponential distribution. The Weibull distribution has a longer and shorter tail than the exponential distribution when $\alpha < 1$ and $\alpha > 1$, respectively.
The Weibull distribution can be expressed as the Laplace transform of a $p(\lambda)$ if and only if
$0<\alpha\le 1$ \cite{Jewell1982AnnStat,Yannaros1994AnnInstStatMath}. The distribution when $\alpha=1/2$ is the so-called stable distribution of order $1/2$, for which we obtain \cite{Feller1971book2,Jewell1982AnnStat,Grandell1997book}
\begin{equation}
p(\lambda) = \frac{m^{\frac{1}{2}}e^{-\frac{m}{4\lambda}}} {2\sqrt{\pi}\lambda^{\frac{3}{2}}}.
\end{equation}

For other values of $\alpha$ (i.e., $0<\alpha<1/2$ or $1/2<\alpha < 1$), the explicit form of $p(\lambda)$ is complicated \cite{Jewell1982AnnStat} such that the use of the Laplace Gillespie algorithm is impractical. 
For these $\alpha$ values, a mixture of a small number of exponential distributions may resemble the Weibull distribution \cite{JinGonigunta2010JStatComputSimul}, such that we may be able to use $p(\lambda)$ with point masses at some discrete values of $\lambda$ to approximate the Weibull distribution of inter-event times.

\item Gamma distribution

When inter-event times obey the gamma distribution, i.e., 
\begin{equation}
\psi(\tau) = \frac{\tau^{\alpha-1}e^{-\tau/\kappa}}{\Gamma(\alpha)\kappa^{\alpha}},
\end{equation}
$\Psi(\tau)$ can be expressed as the Laplace transform of a probability density function $p(\lambda)$ if and only if $0<\alpha\le 1$ \cite{Yannaros1988JApplProb,Gleser1989AmStat}. We obtain \cite{Gleser1989AmStat}
\begin{equation}
p(\lambda) = \begin{cases}
0 & (0 < \lambda < \kappa^{-1}),\\
\frac{1}{\Gamma(\alpha)\Gamma(1-\alpha)\lambda(\kappa\lambda-1)^{\alpha}} & (\lambda\ge \kappa^{-1}).
\end{cases}
\end{equation}

\item Mittag--Leffler distribution

Consider the distribution of inter-event times defined in terms of the survival function given by
\begin{equation}
\Psi(\tau) = E_{\beta}(-\tau^{\beta}),
\label{eq:Psi(tau) Mittag}
\end{equation} 
where
\begin{equation}
E_{\beta}(z) = \sum_{n=0}^{\infty} \frac{z^n}{\Gamma(1+\beta n)}
\label{eq:Mittag-Leffler function}
\end{equation}
is the so-called Mittag--Leffler function. When $0<\beta<1$, $\Psi(\tau)$ is completely monotone, and we obtain
\cite{Gorenflo1997chapter,Gorenflo2014book}
\begin{equation}
p(\lambda) = \frac{1}{\pi} \frac{\lambda^{\beta-1}\sin(\beta\pi)}{\lambda^{2\beta} + 2\lambda^{\beta} \cos(\beta\pi) + 1}.
\end{equation}
When $\beta=1$, Eqs.~\eqref{eq:Psi(tau) Mittag} and \eqref{eq:Mittag-Leffler function} imply that $\Psi(\tau)=e^{-\lambda \tau}$, yielding a Poisson process. When $0<\beta<1$, $\Psi(\tau)$ is long-tailed with the asymptotics \cite{Gorenflo1997chapter,Georgiou2015PhysRevE}
\begin{equation}
\Psi(\tau) \approx \frac{\sin(\beta\pi)\Gamma(\beta)}{\pi \tau^{\beta}},
\end{equation}
or, equivalently,
\begin{equation}
\psi(\tau) \approx \frac{\beta\sin(\beta\pi)\Gamma(\beta)}{\pi \tau^{\beta+1}}.
\end{equation}
Therefore, this class of $\psi(\tau)$ produces long-tailed distributions of inter-event times with a power-law exponent lying between one and two. A special case occurs when $\beta=1/2$, in which case
$\Psi(\tau) = e^{-\tau^{2\beta}}\left[1-\text{erf} (\tau^{\beta})\right]$, where $\text{erf}(z)\equiv (2/\sqrt{\pi})\int_0^z e^{-z^{\prime 2}} {\rm d}z^{\prime}$ is the error function.

\item Integral of a valid survival function

The function given by
\begin{equation}
\Psi^{\rm w}(\tau) \equiv \frac{\int_{\tau}^{\infty}\Psi(\tau^{\prime}){\rm d}\tau^{\prime}}
{\int_0^{\infty}\Psi(\tau^{\prime}){\rm d}\tau^{\prime}}
= \frac{\int_{\tau}^{\infty}\Psi(\tau^{\prime}){\rm d}\tau^{\prime}}
{\langle \tau\rangle_{\psi}}
\label{eq:Psi^w(tau)}
\end{equation}
is well-defined if and only if $\langle \tau\rangle_{\psi}$, i.e., the mean inter-event time with respect to density $\psi(\tau)$ is finite. Assume that the renewal process generated by $\psi(\tau)$ permits use of the Laplace Gillespie algorithm. Because $\Psi^{\rm w}(\tau)\ge 0$ ($\tau\ge 0$), ${\rm d}^n\Psi^{\rm w}(\tau)/{\rm d}\tau^n = - \left[{\rm d}^{n-1}\Psi(\tau)/{\rm d}\tau^{n-1}\right]/\int_0^{\infty}\Psi(\tau^{\prime}){\rm d}\tau^{\prime}$ ($n=1, 2, \ldots$), and $\Psi(\tau)$ is completely monotone, it follows that $\Psi^{\rm w}(\tau)$ is completely monotone. In addition, Eq.~\eqref{eq:Psi^w(tau)} implies that $\Psi^{\rm w}(0)=1$. Therefore, the renewal process with survival function $\Psi^{\rm w}(\tau)$ can also be simulated by the Laplace Gillespie algorithm.

The corresponding probability density of inter-event times is given by
\begin{equation}
\psi^{\rm w}(\tau) = -\frac{{\rm d}\Psi^{\rm w}(\tau)}{{\rm d}\tau} =
\frac{\Psi(\tau)}
{\langle \tau\rangle_{\psi}}.
\label{eq:psi^w(tau)}
\end{equation}
In terms of $p(\lambda)$, we obtain
\begin{equation}
\psi^{\rm w}(\tau) = \frac{\int_0^{\infty} p(\lambda) e^{-\lambda\tau}{\rm d}\lambda}
{\int_0^{\infty}\frac{p(\lambda^{\prime})}{\lambda^{\prime}}{\rm d}\lambda^{\prime}}
\end{equation}
and
\begin{equation}
\Psi^{\rm w}(\tau) = \frac{\int_0^{\infty}\frac{p(\lambda)}{\lambda}e^{-\lambda\tau}{\rm d}\lambda}
{\int_0^{\infty}\frac{p(\lambda^{\prime})}{\lambda^{\prime}}{\rm d}\lambda^{\prime}}.
\end{equation}
Therefore, in each update of the Laplace Gillespie algorithm with the density of inter-event times given by $\psi^{\rm w}(\tau)$, the rate of the Poisson process $\lambda$ should be sampled according to density $p^{\rm w}(\lambda)$, where
\begin{equation}
p^{\rm w}(\lambda) = \frac{\frac{p(\lambda)}{\lambda}}
{\int_0^{\infty}\frac{p(\lambda^{\prime})}{\lambda^{\prime}}{\rm d}\lambda^{\prime}}.
\label{eq:p^w(lambda)}
\end{equation}

For example, if $\psi(\tau)$ is an exponential distribution, then $\psi^{\rm w}(\tau)$ is an exponential distribution of the same mean. If $\psi(\tau)$ is the power-law distribution given by Eq.~\eqref{eq:psi(tau) power law}, then $\psi^{\rm w}(\tau)$ is a power-law distribution of the same form, with $\alpha$ replaced by $\alpha-1$.

\item Product of valid survival functions

The product of two completely monotone functions, $\Psi_1(\tau)$ and $\Psi_2(\tau)$, is completely monotone \cite{Feller1971book2}. In addition, $\Psi_1(0)\Psi_2(0)=1$ if $\Psi_1(0) = \Psi_2(0)=1$. Therefore, the survival function $\Psi(\tau)\equiv \Psi_1(\tau)\Psi_2(\tau)$ admits use of the Laplace Gillespie algorithm if $\Psi_1(\tau)$ and $\Psi_2(\tau)$ do as well. The probability density of the event rate will be the convolution of $p_1(\lambda)$ and $p_2(\lambda)$, where $\Psi_i(\tau)=\int_0^{\infty} p_i(\lambda) e^{-\lambda \tau}{\rm d}\lambda$ ($i=1, 2$).

\item Numerical Laplace transform

Given an arbitrary $p(\lambda)$, we can in principle carry out a numerical Laplace transform to derive $\Psi(\tau)$. We then obtain a valid $\psi(\tau)$ by numerically differentiating $\Psi(\tau)$.

\end{itemize}

\subsection{Empirical distributions of inter-event times}

We are often interested in informing multivariate point processes by empirical data of event sequences. A standard numerical approach is to emulate the dynamics (e.g., epidemic processes) on top of empirical event sequences, i.e., use empirically observed events with time stamps to induce, for example, infection events \cite{HolmeSaramaki2012PhysRep,Masuda2013F1000}. There exists a Gillespie algorithm to run dynamical processes on such empirical temporal networks \cite{Vestergaard2015PlosComputBiol}. Another approach is to estimate $\psi(\tau)$ from empirical data and then use a variant of the Gillespie algorithm (e.g., the nMGA or the Laplace Gillespie algorithm) to simulate stochastic point processes (e.g., epidemic processes).

The Laplace Gillespie algorithm requires the survival function, $\Psi(\tau)$, to be completely monotone. Under this condition, we may be able to compute the inverse Laplace transform to obtain $p(\lambda)$ at a reasonable computational cost \cite{Abate1992QueueSyst}. However, because it is likely that an empirical $\Psi(\tau)$ is not completely monotone, we propose two alternative methods to estimate $p(\lambda)$ from empirical data. The first method is to fit a completely monotone survival function of inter-event times, such as Eq.~\eqref{eq:Psi(tau) power law}, to given data. The second method is to estimate a finite mixture of exponential distributions of different means to approximate the empirical $\psi(\tau)$ or $\Psi(\tau)$. Likelihood or other cost-function methods are available for performing this estimation \cite{Jewell1982AnnStat,Heckman1990JAmStatAssoc,Kleinberg2003DataMinKnowlDisc,Politi2007PhysicaA,JinGonigunta2010JStatComputSimul}. If the empirical $\Psi(\tau)$ is completely monotone, the approximation error is guaranteed to decay inversely proportional to the number of constituent distributions \cite{LiBarron2000NIPS}. For both of these methods, we should be mindful of the bias in the estimation caused by a finite time window of observation \cite{Kivela2015PhysRevE}.

\subsection{Initial conditions}

When we begin to run $N$ processes, one approach is to initially draw the inter-event time for each process from $\psi(\tau)$. This initial condition defines the so-called ordinary renewal process \cite{Cox1962book}. An alternative model, called the equilibrium renewal process, assumes that the process has begun at time $t=-\infty$, such that the first inter-event time for each process, drawn at $t=0$, uses the equilibrium distribution of waiting times to the next event, rather than $\psi(\tau)$ (i.e., the distribution of inter-event times) \cite{Cox1962book}. In fact, the equilibrium distribution of waiting times to the next event coincides with $\psi^{\rm w}(\tau)$ given by Eq.~\eqref{eq:psi^w(tau)} \cite{Cox1962book,Feller1971book2,Masuda2016book}. To simulate the equilibrium renewal process, we start by drawing the rates of the Poisson processes according to $p^{\rm w}(\lambda)$ given by Eq.~\eqref{eq:p^w(lambda)}. Afterwards, we draw the event rates according to $p(\lambda)$.

\section{Numerical performance}\label{sec:numerical}

In this section, we compare the performances of the nMGA and the Laplace Gillespie algorithm.
We use the power-law distribution of inter-event times given by Eq.~\eqref{eq:psi(tau) power law}.
Because $\kappa$ only controls the scale of inter-event times, we set $\kappa=1$ without loss of generality.
To generate gamma-distributed random variates, we use a well-known algorithm \cite{Marsaglia2000ACMTransMathSoft} and adapt an open source code \cite{Marsaglia-code} for our purposes. We generate $\lfloor N/3 \rfloor$ processes by Eq.~\eqref{eq:psi(tau) power law} with $\alpha=1$, another $\lfloor N/3 \rfloor$ processes with $\alpha=1.5$, and another $N-2\lfloor N/3\rfloor \approx N/3$ processes with $\alpha=2$. We continue the simulation until one of the $N$ processes generates $10^6$ inter-event times. We employ the ordinary renewal process such that the initial inter-event time for each process is drawn from $\psi(\tau)$.

\begin{figure}
\begin{center}
\includegraphics[width=8cm]{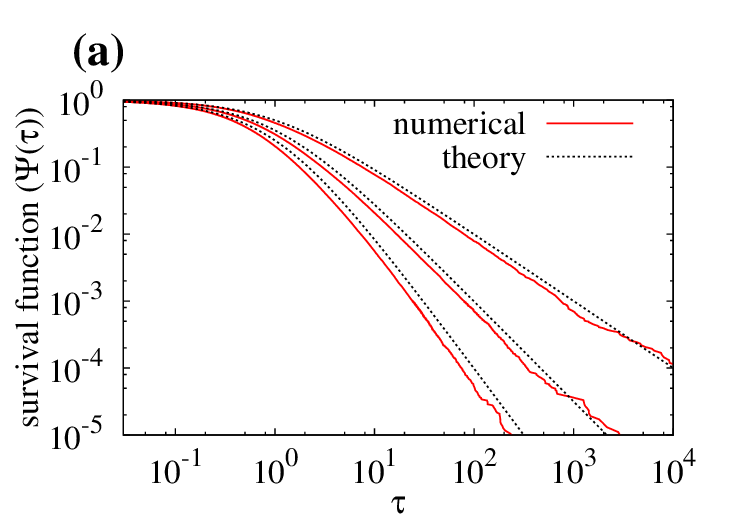}
\includegraphics[width=8cm]{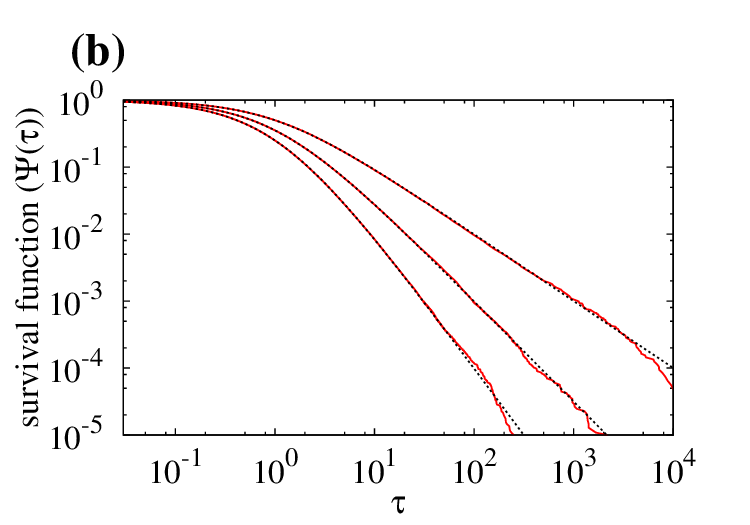}
\includegraphics[width=8cm]{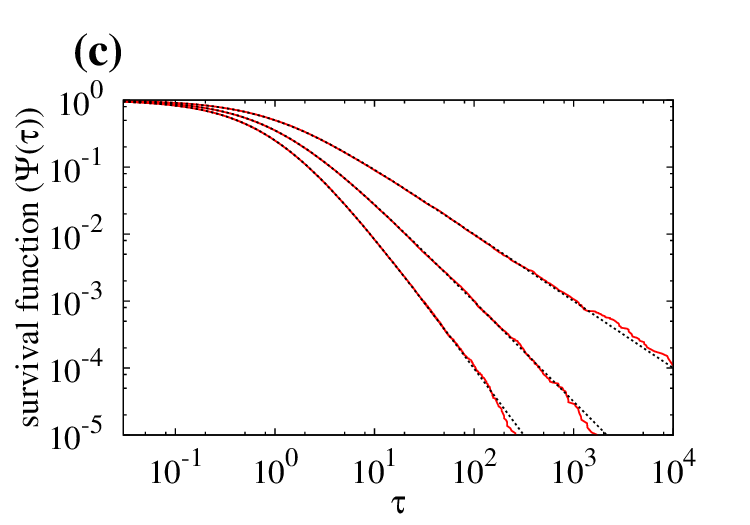}
\includegraphics[width=8cm]{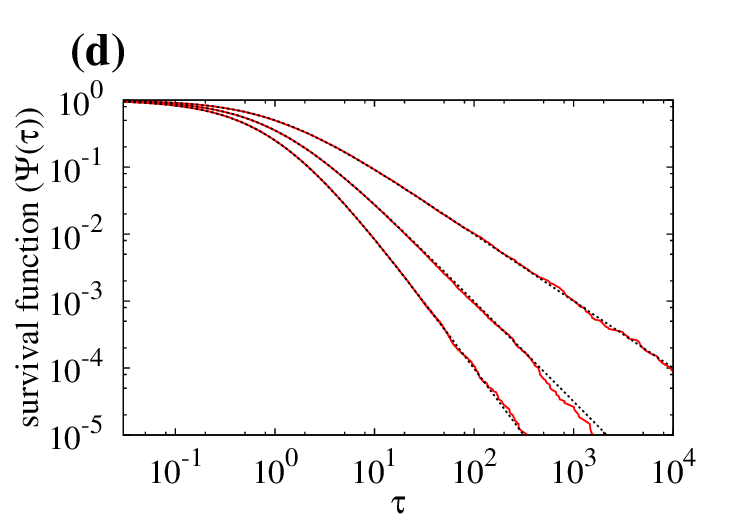}
\includegraphics[width=8cm]{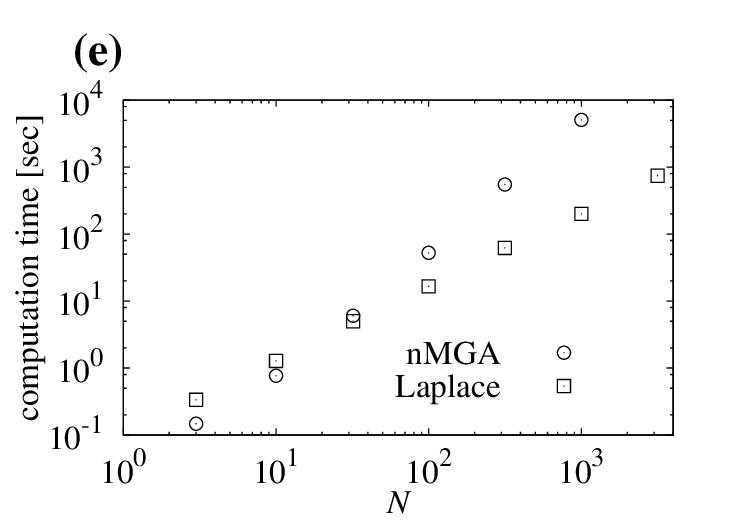}
\caption{Comparison between the nMGA and the Laplace Gillespie algorithm. The distribution of inter-event times is power law (Eq.~\eqref{eq:psi(tau) power law}) with $\kappa=1$. Among the $N$ processes, $\lfloor N/3 \rfloor$ processes are simulated with $\alpha=1$, another $\lfloor N/3 \rfloor$ processes with $\alpha=1.5$, and another $N-2\lfloor N/3 \rfloor \approx$ $N/3$ processes with $\alpha=2$. (a)--(d) Survival function of inter-event times for one process with $\alpha=1$, another with $\alpha=1.5$, and another with $\alpha=2$, from the top to the bottom. (a) nMGA when $N=10$. (b) Laplace Gillespie algorithm when $N=10$. (c) nMGA when $N=10^3$. (d) Laplace Gillespie algorithm when $N=10^3$. (e) Computation time as a function of the number of processes, $N$.}
\label{fig:survival}
\end{center}
\end{figure}

The survival functions for one of the processes with $\alpha=1$, one of those with $\alpha=1.5$, and one of those with $\alpha=2$ are shown by the solid curves for the nMGA and the Laplace Gillespie algorithm in Figs.~\ref{fig:survival}(a) and \ref{fig:survival}(b), respectively, for $N=10$. The theoretical survival function, Eq.~\eqref{eq:Psi(tau) power law}, is plotted with the dashed curves. The results obtained from the Laplace Gillespie algorithm (Fig.~\ref{fig:survival}(b)) are more accurate than those obtained from the nMGA. This is because the nMGA is exact only in the limit of $N\to\infty$, whereas the Laplace Gillespie algorithm is exact for any $N$. When $N=10^3$, the nMGA is sufficiently accurate (Fig.~\ref{fig:survival}(c)), as is the Laplace Gillespie algorithm (Fig.~\ref{fig:survival}(d)). The results shown in Figs.~\ref{fig:survival}(a) and \ref{fig:survival}(c) are consistent with the numerical results obtained in Ref.~\cite{BogunaLafuerza2014PhysRevE}.

The nMGA may require a lot of time in updating the instantaneous event rates for all processes every time an event occurs in one of the $N$ processes. The Laplace Gillespie algorithm avoids this rate recalculation, whereas it might be costly to calculate the gamma variates each time an event occurs. We compare numerically the computation times for the two algorithms by varying $N$. The other parameters remain the same as those used in Figs.~\ref{fig:survival}(a)--(d). For the Laplace Gillespie algorithm, we use a binary tree data structure to store and update $\lambda_i$ ($1\le i\le N$) to accelerate the selection of the $i$ value with probability $\Pi_i$ upon the occurrence of each event \cite{GibsonBruck2000JPhysChemA}. In short, each $\lambda_i$ occupies a leaf of the tree, and each non-leaf node stores the sum of its left child and its right child. This data structure is useful when only a small number of $\lambda_i$ are changed following the occurrence of each event \cite{GibsonBruck2000JPhysChemA}. This is not the case for the nMGA, for which all the $N$ instantaneous event rates must be updated upon each event. Therefore, for the nMGA, we use a simple linear search, which is computationally less expensive than updating a binary tree every time an event occurs. We use codes written in C++, compiled with a standard g++ compiler without an optimisation option on a Mac Book Air with 1.7 GHz Intel Core i7 processor and 8GB 1600 MHz DDR3 RAM. The computation time in seconds plotted as a function of $N$ in Fig.~\ref{fig:survival}(e) indicates that the Laplace Gillespie algorithm runs substantially faster than the nMGA as $N$ increases.

Both the nMGA and the Laplace Gillespie algorithm require two uniformly distributed random variates per event, as does the standard Gillespie algorithm. In addition, for each generated event, the nMGA demands $O(N)$ time to search for the process to fire and update the instantaneous event rates for all processes. In contrast, the Laplace Gillespie algorithm demands $O(k \log N)$ time per event on average, where $k$ is the typical number of processes that are affected by the firing of the $i$th process. The search for what process to fire requires $O(\log N)$ time, given the binary tree data structure \cite{GibsonBruck2000JPhysChemA}. The updating of $\lambda_j$ due to the event of the $i$th process requires $O(k)$ time, including the generation of $k$ gamma-distributed random variates. The updating of the binary tree requires $O(k\log N)$ time because it consumes $O(\log N)$ time for each updated $\lambda_j$. Because $k$ is usually much smaller than $N$ in practice \cite{GibsonBruck2000JPhysChemA,HolmeSaramaki2012PhysRep}, the Laplace Gillespie algorithm is expected to run faster than the nMGA, which is consistent with the numerical results shown in Fig.~\ref{fig:survival}(e).

Additionally, the Laplace Gillespie algorithm outperforms the nMGA in the sense that the Laplace Gillespie algorithm is exact for any $N$, whereas the nMGA is not. On the other hand, for the Laplace Gillespie algorithm, the form of $\psi(\tau)$ is limited, whereas the nMGA allows for any $\psi(\tau)$ to be used.

\section{Positively correlated sequences of inter-event times\label{sub:positive correlation}}

We have considered renewal processes, i.e., stationary point processes without correlation between inter-event times. However, inter-event times are positively correlated for human activity and earthquakes \cite{GohBarabasi2008EPL,Masuda2013hawkes}. The Laplace Gillespie algorithm provides a method for generating point processes with positive correlation, without changing $\psi(\tau)$. To generate such event sequences, we redraw a new event rate for the Poisson process, $\lambda_i$, with probability $1-q$ ($0\le q< 1$), when the $i$th process has generated an event. With probability $q$, we continue to use the same value of $\lambda_i$ until the $i$th process generates another event. We call this algorithm the correlated Laplace Gillespie algorithm. The standard Laplace Gillespie algorithm is recovered when $q=0$. The correlation between inter-event times grows as $q$ increases. Although the same $\lambda_i$ value may be used for generating consecutive inter-event times, the corresponding inter-event times are different because they are generated from a Poisson process. The computation time for the correlated Laplace Gillespie algorithm decreases as $q$ increases because the number of times that $\lambda_i$ is redrawn is proportional to $1-q$.

In a continuous-time Markov process with a state-dependent hopping rate, the inter-event time defined as the time between two consecutive hops, regardless of the state, is generally correlated across inter-event times \cite{Schwalger2010EurPhysJSpecTopics}. The correlated Laplace Gillespie algorithm can be interpreted as a special case of this scenario such that the state is continuous, the process hops back to the current state with probability $q$, and it hops to any other state with probability proportional to $(1-q)\times p(\lambda)$. The correlated Laplace Gillespie algorithm can be alternatively built on top of a finite-state \cite{Schwalger2010EurPhysJSpecTopics} or an infinite-state \cite{Kleinberg2003DataMinKnowlDisc} Markov process with a general transition probability between states. This variant of the correlated Laplace Gillespie algorithm is similar to a two-state cascading Poisson process in which the two states correspond to different event rates \cite{Malmgren2008PNAS}.

Two remarks are now in order. First, $\psi(\tau)$ is independent of the $q$ value. This is because the stationary density of the corresponding continuous-time Markov process in the $\lambda$-space is equal to $p(\lambda)$, irrespective of the $q$ value. Of course, the distribution of $\tau$ conditioned on the previous inter-event time, $\tau^{\prime}$, is different from $\psi(\tau)$ and depends on $\tau^{\prime}$ in general. Second, the correlated Laplace Gillespie algorithm cannot be used to generate correlated event sequences when $\psi(\tau)$ is the exponential distribution. In this case, the event rate $\lambda$ must be kept constant over time and therefore cannot be modulated in a temporally correlated manner.

We measure the so-called memory coefficient \cite{GohBarabasi2008EPL} to quantify the amount of correlation in a sequence of inter-event times generated by the correlated Laplace Gillespie algorithm. The memory coefficient for a sequence of inter-event times, $\{\tau_1, \tau_2, \ldots, \tau_n\}$, where $n$ is the number of inter-event times, is defined by
\begin{equation}
M = \frac{\sum_{i=1}^{n-1} (\tau_i-m_1)(\tau_{i+1}-m_2)}
{\sqrt{\sum_{i=1}^{n-1} (\tau_i-m_1)^2 \sum_{i=2}^n (\tau_{i+1}-m_2)^2}},
\end{equation}
where $m_1 = \sum_{i=1}^{n-1} \tau_i / (n-1)$ and $m_2 = \sum_{i=2}^n \tau_i/(n-1)$.

\begin{figure}
\begin{center}
\includegraphics[width=8cm]{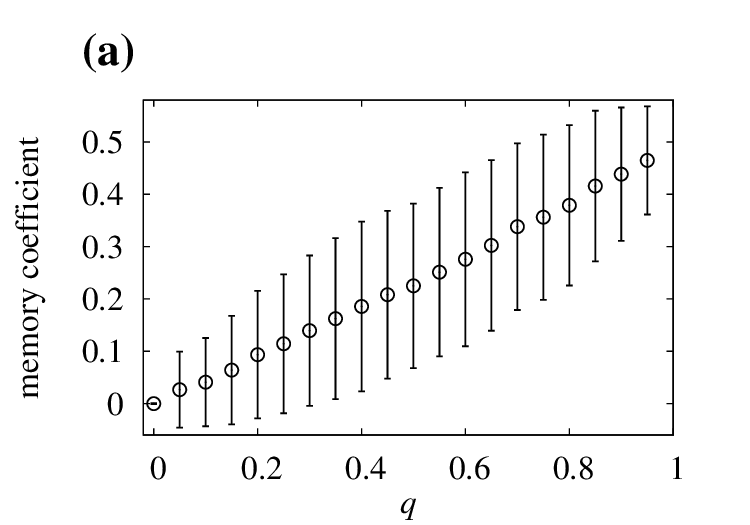}
\includegraphics[width=8cm]{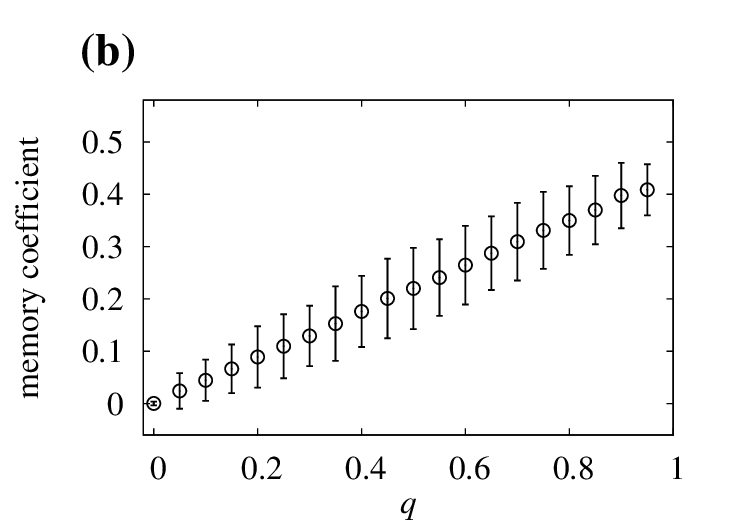}
\includegraphics[width=8cm]{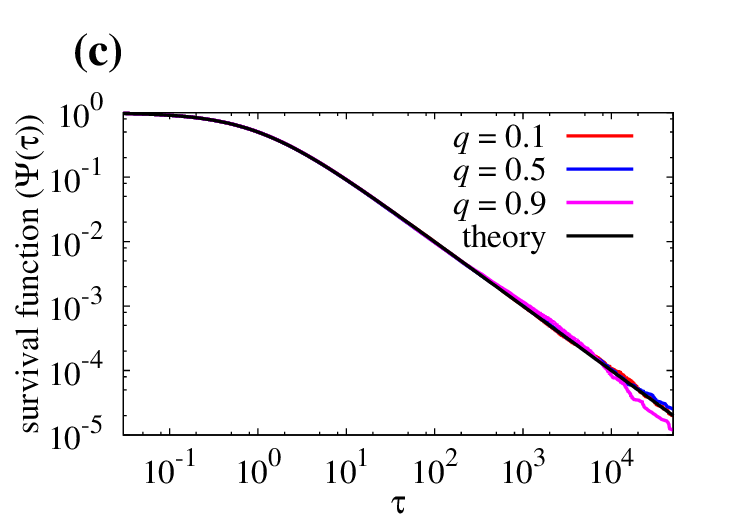}
\includegraphics[width=8cm]{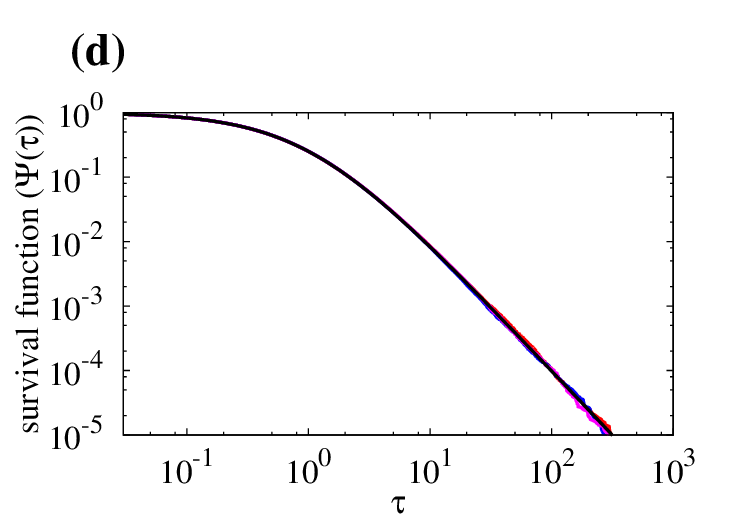}
\caption{Memory coefficient, $M$, for the correlated Laplace Gillespie algorithm.
We used the power-law distribution of inter-event times given by Eq.~\eqref{eq:psi(tau) power law} with $\kappa=1$. (a) $\alpha=1$. (b) $\alpha=2$. The error bar represents the mean $\pm$ standard deviation. (c) Survival function of a single event sequence (i.e., $N=1$) with $10^6$ events with $\alpha=1$ and $q=0.1$, 0.5, and 0.9. (d) Similar to (c) with $\alpha=2$.}
\label{fig:correlated}
\end{center}
\end{figure}

For the power-law distribution of inter-event times given by Eq.~\eqref{eq:psi(tau) power law} with $\kappa=1$, we generate a sequence of $n=10^5$ inter-event times and calculate $M$. The mean and standard deviation of $M$, calculated on the basis of $10^3$ sequences, are plotted for $\alpha=1$ and $\alpha=2$ in Figs.~\ref{fig:correlated}(a) and \ref{fig:correlated}(b), respectively. For both $\alpha$ values, $M$ monotonically increases with $q$ and a range of $M$ values between 0 and $\approx 0.4$ is produced. In empirical data, $M$ lies between 0 and 0.1 for human activity and between 0.1 and 0.25 for natural phenomena \cite{GohBarabasi2008EPL}. These ranges of $M$ are produced using approximately $0\le q\le 0.2$ and $0.2 \le q\le 0.5$, respectively.

\section{Epidemic processes}\label{sec:epidemic}

Previous numerical efforts suggested that the dynamics of epidemic processes in well-mixed populations or networks were altered if contact events were generated by non-Poissonian renewal processes
\cite{Rocha2013PlosComputBiol,Vanmieghem2013PhysRevLett,MinGohKim2013EPL,Horvath2014NewJPhys}. The nMGA and the Laplace Gillespie algorithm can be used for implementing such models of epidemic processes. To demonstrate the use of the Laplace Gillespie algorithm, we simulate a node-centric susceptible-infected-recovered (SIR) epidemic process model, which is similar to previous models \cite{Rocha2011PlosComputBiol,Rocha2013PlosComputBiol,Perra2012SciRep}.

Consider a static network composed of $N$ nodes. At any point in time, each node assumes one of the three states: susceptible, infected, or recovered. An infected node $i$ transmits the disease to a susceptible node $j$ upon the activation of link $(i, j)$. To activate links, we initially assign to each node $i$ ($1\le i\le N$) an independent and identical point process whose probability density of inter-event times is given by $\psi(\tau)$. When an event occurs at node $i$, we select a neighbour of $i$, denoted by $j$, with the equal probability and activate link $(i, j)$. If either $i$ or $j$ is infected and the other is susceptible, the disease is transmitted such that the susceptible node becomes infected. An infected node transits to the recovered state according to a Poisson process of rate $\mu$. A recovered node neither infects nor is infected by other nodes.

The mean time to node activation, which enables infection, is given by $\langle \tau\rangle = \int_0^{\infty}\tau \psi(\tau){\rm d}\tau$. The mean time for an infected node to recover is equal to $1/\mu$. We define the effective infection rate by $\lambda_{\rm eff} = (1/\mu) / \langle \tau\rangle$
\cite{BogunaLafuerza2014PhysRevE}. We control $\lambda_{\rm eff}$ by changing $\mu$ for a given $\psi(\tau)$. This definition is justified because multiplying $\langle \tau\rangle$ and $1/\mu$ by the same factor only changes the time scale of the dynamics.

We assume an equilibrium point process, i.e., we start simulations from the equilibrium state. This is equivalent to drawing the first event time for each node from the waiting-time distribution, $\psi^{\rm w}(\tau)$, rather than from $\psi(\tau)$, and drawing subsequent event times from $\psi(\tau)$. The population structure is assumed to be either well mixed (i.e., complete graph) or the regular random graph in which all nodes have degree five and all links are randomly formed. In both cases, we set $N=10^4$. Each simulation starts from the same initial condition, in which a particular node, which is the same in all simulations, is infected and all the other $N-1$ nodes are susceptible. We measure the number of recovered nodes at the end of the simulation normalised by $N$, called the final size, averaged over $10^4$ simulations. We consider four renewal processes for node activation: the exponential distribution, corresponding to a Poisson process, and three power-law distributions given by Eq.~\eqref{eq:psi(tau) power law} with $\alpha=1.5$, $\kappa=1$, and $q=0$, 0.2, and 0.9.

\begin{figure}
\begin{center}
\includegraphics[width=8cm]{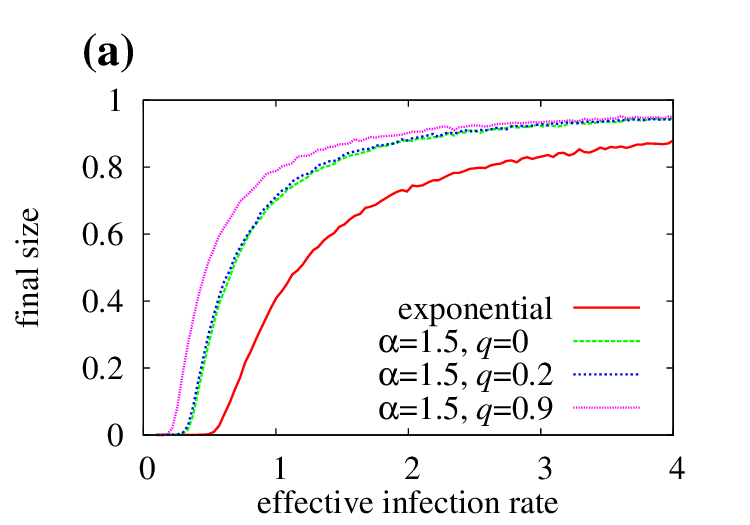}
\includegraphics[width=8cm]{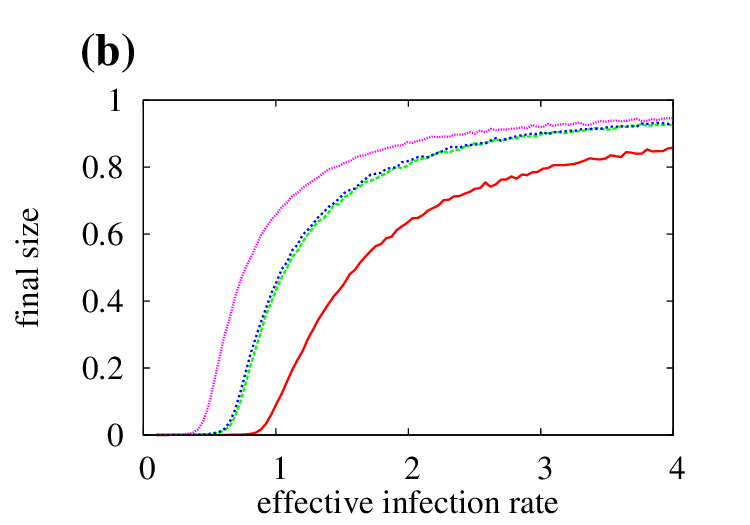}
\caption{Final outbreak size for the SIR epidemic model. (a) Well-mixed population. (b) Regular random graph with degree of each node equal to five. We set $N=10^4$. For the power-law density of inter-event times, we use Eq.~\eqref{eq:psi(tau) power law} with $\kappa=1$ and $\alpha=1.5$.}
\label{fig:epidemic}
\end{center}
\end{figure}

The final size for the well-mixed population and the regular random graph is shown in Figs.~\ref{fig:epidemic}(a) and \ref{fig:epidemic}(b), respectively. For both population structures, and across the entire range of the effective infection rate, $\lambda_{\rm eff}$, the final size is larger when $\psi(\tau)$ is the power-law distribution than when it is the exponential distribution. Consistent with this result, the epidemic threshold, i.e., the value of $\lambda_{\rm eff}$ at which the final size becomes positive, is smaller for the power-law distribution $\psi(\tau)$ than for the exponential distribution $\psi(\tau)$.

The final size is larger with positive correlation of inter-event times ($q=0.9$) than with no correlation ($q=0$). Results for $q=0.2$ are almost identical to those for $q=0$. Because realistic values of the memory coefficient, $M$, for human activity are produced by $0\le q\le 0.2$ (section~\ref{sub:positive correlation}), we conclude that a realistic amount of positive correlation in inter-event times does not affect the final size.

\section{Discussion}\label{sec:discussion}

We have provided a generalisation of the Gillespie algorithm for non-Poissonian renewal processes, which we call the Laplace Gillespie algorithm. Our algorithm is exact for any number of processes running in parallel and is faster than the nMGA \cite{BogunaLafuerza2014PhysRevE}. Although it is only applicable to renewal processes whose survival function is completely monotone, it applies to several renewal processes of interest. We have also proposed a method to simulate non-renewal point processes with tunable positive correlation between inter-event times.

Previous studies numerically explored Poissonian explanations of long-tailed distributions of inter-event times. Examples include a non-homogeneous Poisson process whose event rate switches between two values and is also periodically modulated \cite{Malmgren2008PNAS}. Another example is a self-exciting Hawkes process with an exponential memory kernel \cite{Masuda2013hawkes}. We have shown that a power-law distribution of inter-event times, Eq.~\eqref{eq:psi(tau) power law}, is generated when the rate of an event-modulated Poisson process is drawn from the gamma distribution upon the occurrence of every event. This observation provides a theoretical underpinning of the fact that non-homogeneous Poisson processes and Hawkes processes can generate long-tailed distributions of inter-event times. In other words, switching between different rates is a general mechanism to produce long-tailed distributions of inter-event times. Although the present results indicate that, in theory, we require a mixture of an infinite number of Poisson processes of different rates to produce a power-law distribution, in practice a small number of Poisson processes may be sufficient. In fact, a mixture of a small number of exponential distributions is sometimes employed to fit empirical distributions of inter-event times \cite{Jewell1982AnnStat,Heckman1990JAmStatAssoc,Politi2007PhysicaA,JinGonigunta2010JStatComputSimul}.

We have applied the Laplace Gillespie algorithm to an epidemic model in well-mixed and networked populations. The applicability of the Laplace Gillespie algorithm, as well as of the modified next reaction method \cite{Anderson2007JChemPhys} and of the nMGA, extends far beyond epidemic modelling.
In fact, these algorithms can simulate systems of earthquakes, spiking neurons, financial time series, crimes, and so on (see section~\ref{sec:introduction} for references). In particular, empirical data corresponding to these applications suggest long-tailed distributions of inter-event times (section~\ref{sec:introduction}), thus yielding a CV (coefficient of variation, i.e., the standard deviation divided by the mean) larger than unity, and therefore not excluding the use of the Laplace Gillespie algorithm. It is also straightforward to include births and deaths of nodes  \cite{Bansal2010JBiolDyn,Rocha2013PlosComputBiol} and links \cite{Holme2014SciRep} of contact networks, or rewiring of links \cite{Volz2007ProcRSocB,GrossBlasius2008JRSocInterface}, as long as these events obey renewal processes or non-renewal point processes with positive correlation, as emulated by the correlated Laplace Gillespie algorithm.

The Laplace Gillespie algorithm can be employed if and only if the survival function of inter-event times is completely monotone. Some convenient conditions for and examples of survival functions that are not completely monotone are as follows:

(i) Non-monotonicity. By setting $n=2$ in Eq.~\eqref{eq:completely monotone}, we obtain ${\rm d\psi(\tau)}/{\rm d}\tau \le 0$. Therefore, $\psi(\tau)$ must monotonically decrease with $\tau$ for the Laplace Gillespie algorithm to be applicable. This condition excludes the gamma and Weibull distributions with shape parameter $\alpha>1$, any log-normal distribution, and any Pareto distribution, i.e.,
\begin{equation}
\psi(\tau)= 
\begin{cases}
\frac{\alpha}{\tau_0}\left(\frac{\tau_0}{\tau}\right)^{\alpha+1} & (\tau\ge \tau_0),\\
0 & (\tau < \tau_0),
\end{cases}
\end{equation}
where $\alpha>0$ and $\tau_0>0$.

(ii) CV is smaller than unity. Complete monotonicity implies that the CV of $\tau$ is larger than or equal to unity \cite{Yannaros1994AnnInstStatMath}. This condition again excludes the gamma and Weibull distributions with $\alpha>1$. In practice, a CV value less than unity indicates that events occur more regularly than in a Poisson process, which would yield $\text{CV} = 1$. Therefore, renewal processes producing relatively periodic event sequences are also excluded.

In epidemiology, evidence suggests that empirical recovery times are less dispersed than the exponential distribution, implying a CV value less than unity. Therefore, a gamma distribution with scale parameter $\alpha>1$ or even a delta distribution is often alternatively employed 
\cite{Lloyd2001ProcRSocLondB,Wearing2005PlosMed}. These distributions cannot be simulated by our algorithm.

(iii) Higher-order conditions. Even when ${\rm d}\psi(\tau)/{\rm d}\tau \le 0$ and the CV is large, $\Psi(\tau)$ may not be completely monotone. For example, the one-sided Cauchy distribution defined by $\psi(\tau) = 1/\left[\pi(\tau^2 + 1)\right]$ yields ${\rm d}^2\psi(\tau)/{\rm d}\tau^2 = 2(3\tau^2-1)/\left[\pi(\tau^2+1)^3\right]$, whose sign depends on the value of $\tau$.

Empirical evidence of online correspondences of humans suggests that, except for very small $\tau$ values, $\tau$ obeys a power-law distribution for small $\tau$ and an exponential distribution for large $\tau$ \cite{WuZhou2010PNAS}. Such a $\psi(\tau)$ monotonically decreases with $\tau$, verifying that ${\rm d}\psi(\tau)/{\rm d}\tau \le 0$. However, the sign of ${\rm d}^2\psi(\tau)/{\rm d}\tau^2$ depends on the $\tau$ value, such that the corresponding survival function is not completely monotone.
 
\section*{Appendix A: Halving the number of random variates used in the Gillespie algorithm}

Each step of the Gillespie algorithm usually requires two random variates, $u_1$ and $u_2$, uniformly distributed on $[0, 1]$, one to draw the time increment via $\Delta t = -\log u_1\big/ \left(\sum_{i=1}^N \lambda_i\right)$ and the other to select the process $i$ that fires on the basis of Eq.~\eqref{eq:Pi(i) Gillespie}. To eliminate one random variate, we first select $i$ using random variate $u_2$. Independently of whether we use a binary or a linear search, we end up identifying the unique $i$ value that satisfies $\sum_{j=1}^{i-1} \Pi_j \le u_2 < \sum_{j=1}^i \Pi_j$. Once $i$ is determined, we set $u_1\equiv (u_2 - \sum_{j=1}^{i-1} \Pi_j)/\Pi_i$, which is in fact uniformly distributed on $[0, 1]$. Therefore, we do not have to draw $u_1$ using a random number generator. This mathematical trick is similar to the one employed in the next reaction methods \cite{GibsonBruck2000JPhysChemA,Anderson2007JChemPhys}.

\section*{Appendix B: Derivation of the exact Gillespie algorithm for general renewal processes}

Consider $N$ renewal processes running in parallel. If the $i$th process is running in isolation, the waiting time $\tau$ until the next event is distributed according to
\begin{equation}
\psi_i^{\rm w}(\tau | t_i) = \frac{\psi_i(t_i+\tau)}{\Psi_i(t_i)},
\label{eq:psi(tau | t_i) generalized Gillespie}
\end{equation}
where $\Psi_i(t_i)$ is the survival function of the $i$th process given by Eq.~\eqref{eq:Psi_i(t_i)}.

In fact, the $i$th process coexists with the other $N-1$ processes. We denote by $\phi(\Delta t, i | \{t_j\})$ the probability density with which the $i$th process, but not the other $N-1$ processes, generates the next event in the set of $N$ processes after time $\Delta t$, given the time since the previous event for each process, $\{t_j\}$, i.e., $t_1, \ldots, t_N$. We obtain
\begin{equation}
\phi(\Delta t, i | \{t_j\}) = \psi_i^{\rm w}(\Delta t | t_i) \prod_{j=1; j\neq i}^N \Psi_j(\Delta t | t_j),
\label{eq:phi(Delta t, i | set of t_j) Gillespie}
\end{equation}
where
\begin{equation}
\Psi_j(\Delta t | t_j) = \int_{\Delta t}^{\infty} \psi_j^{\rm w}(\tau^{\prime} | t_j) {\rm d}\tau^{\prime} =
\frac{\Psi_j(t_j+\Delta t)}{\Psi_j(t_j)}
\label{eq:Psi_j(Delta t | t_j) Gillespie}
\end{equation}
is the probability that the time until the next event for the hypothetically isolated $j$th process is larger than $\Delta t$, conditioned on the assumption that the last event has occurred time $t_j$ before. Using Eqs.~\eqref{eq:psi(tau | t_i) generalized Gillespie} and \eqref{eq:Psi_j(Delta t | t_j) Gillespie}, we rewrite Eq.~\eqref{eq:phi(Delta t, i | set of t_j) Gillespie} as
\begin{equation}
\phi(\Delta t, i | \{t_j\}) = \frac{\psi_i(t_i+\Delta t)}{\Psi_i(t_i+\Delta t)} \Phi(\Delta t | \{t_j\}),
\label{eq:phi(Delta t, i | set of t_j) Gillespie 2}
\end{equation}
where $\Phi(\Delta t | \{t_j\})$ is given by Eq.~\eqref{eq:Phi(Delta t | set of t_j) Gillespie}.

Equation~\eqref{eq:Phi(Delta t | set of t_j) Gillespie} represents the probability that no process generates an event for time $\Delta t$. By equating this quantity to $u$, a random variate over the unit interval, we can determine $\Delta t$, i.e., the time until the next event in the entire population of the $N$ renewal processes. Equation~\eqref{eq:phi(Delta t, i | set of t_j) Gillespie 2} implies that, once $\Delta t$ is determined, $\lambda_i(t_i+\Delta t) = \psi_i(t_i+\Delta t) / \Psi_i(t_i+\Delta t)$ is the instantaneous rate of the $i$th process and is proportional to the probability that the $i$th process generates this event. 
Therefore, the exact Gillespie algorithm for general renewal processes is produced as given in section~\ref{sec:nMGA}.

\section*{Acknowledgments}

N.M. acknowledges the support provided through JST, CREST, and JST, ERATO, Kawarabayashi Large Graph Project. L.E.C.R. is a Charg\'{e} de recherche of the Fonds de la Recherche Scientifique - FNRS.


\newpage
\clearpage

\newpage
\clearpage

\newpage
\clearpage

\newpage
\clearpage

\begin{table}
\begin{center}
\caption{Distributions of inter-event times, $\psi(\tau)$, for which the Laplace Gillespie algorithm can be used. $H$ is the Heaviside function defined by $H(x) = 1$ ($x\ge 0$) and $H(x)=0$ ($x<0$).}
\label{tab:examples}
\begin{tabular}{|c|c|c|c|c|}\hline
Distribution & $\psi(\tau)$ & $\Psi(\tau)$ & condition & $p(\lambda) (\lambda\ge 0)$ \\ \hline\hline
Exponential &  $\lambda_0 e^{-\lambda_0 \tau}$ & $e^{-\lambda_0 \tau}$ &  & $\delta(\lambda-\lambda_0)$ \rule[0mm]{0mm}{8mm}\\ \hline 
Power law & $\frac{\lambda^{\alpha-1}e^{-\lambda/\kappa}}{\Gamma(\alpha)\kappa^{\alpha}}$ &
$\frac{\kappa}{(1+\kappa\tau)^{\alpha+1}}$ &  & $\frac{1}{(1+\kappa\tau)^{\alpha}}$ \rule[0mm]{0mm}{8mm}\\ \hline
Power law with & \multirow{2}{*}{$\frac{e^{-\lambda_0 \tau}}{(1+\kappa\tau)^{\alpha}}\left(\lambda_0+\frac{\kappa\alpha}{1+\kappa\tau}\right)$} & \multirow{2}{*}{$\frac{e^{-\lambda_0 \tau}}{(1+\kappa\tau)^{\alpha}}$} & \multirow{2}{*}{} & \multirow{2}{*}{
$\frac{(\lambda-\lambda_0)^{\alpha-1}e^{-\frac{\lambda-\lambda_0}{\kappa}}H(\lambda-\lambda_0)}{\Gamma(\alpha)}
$} \rule[0mm]{0mm}{4mm}\\
exponential tail &&&& \rule[0mm]{0mm}{4mm}\\ \hline
Power law (with & \multirow{2}{*}{$\propto \frac{e^{-\lambda_{\min}\tau}}{\tau}$ or $\propto \frac{1}{\tau^2}$} & \multirow{2}{*}{$\frac{e^{-\lambda_{\min}\tau}-e^{-\lambda_{\max}\tau}}{\tau\left(\lambda_{\max}-\lambda_{\min}\right)}$} & \multirow{2}{*}{$\lambda_{\min}\ll \lambda_{\max}$} & \multirow{2}{*}{uniform on $[\lambda_{\min}, \lambda_{\max}]$} \rule[0mm]{0mm}{4mm}\\ 
exponential tail) &&&& \rule[0mm]{0mm}{4mm}\\ \hline
Power law & $\frac{\alpha+1}{\tau^{\alpha+2}} \gamma(\alpha+2,\tau)$ & $\frac{\alpha+1}{\tau^{\alpha+1}} \gamma(\alpha+1,\tau)$ & $\alpha>-1$ & $(\alpha+1) \lambda^{\alpha}$ \rule[0mm]{0mm}{8mm}\\ \hline
Weibull & $\alpha\mu^{\alpha}\tau^{\alpha-1} e^{-(\mu\tau)^{\alpha}}$ & $e^{-(\mu\tau)^{\alpha}}$
& $0<\alpha\le 1$ & complicated \rule[0mm]{0mm}{8mm}\\ \hline
Gamma & $\frac{\tau^{\alpha-1}e^{-\tau/\kappa}}{\Gamma(\alpha)\kappa^{\alpha}}$ & complicated & 
$0<\alpha\le 1$ & $\frac{H(\lambda-\kappa^{-1})}{\Gamma(\alpha)\Gamma(1-\alpha)\lambda(\kappa\lambda-1)^{\alpha}}$ \rule[0mm]{0mm}{8mm}\\ \hline
Mittag--Leffler & $\approx \frac{\beta\sin(\beta\pi)\Gamma(\beta)}{\pi \tau^{\beta+1}}$ & $E_{\beta}(-\tau^{\beta})$
& $0<\beta<1$ & $\frac{1}{\pi} \frac{\lambda^{\beta-1}\sin(\beta\pi)}{\lambda^{2\beta} + 2\lambda^{\beta} \cos(\beta\pi) + 1}$ \rule[0mm]{0mm}{8mm}\\ \hline
\end{tabular}
\end{center}
\end{table}

\end{document}